\documentclass{article}
\title{Project Report: On Cryptanalysis via Approximation of Cryptographic 
Primitives Relying on the Planted Clique Conjecture}
\author{Aubrey Alston (ada2145@columbia.edu), Yanrong Wo (yw2513@columbia.edu)}

\usepackage[left=2.5cm,top=3cm,right=2.5cm,bottom=3cm,bindingoffset=0.5cm]{geometry}

\makeatletter
\renewcommand\paragraph{\@startsection{paragraph}{4}{\z@}%
            {-2.5ex\@plus -1ex \@minus -.25ex}%
            {1.25ex \@plus .25ex}%
            {\normalfont\normalsize\bfseries}}
\makeatother
\usepackage{fancyhdr}
\date{}

\usepackage{mathtools}
\usepackage{amsmath}
\usepackage{amssymb}
\usepackage{MnSymbol}
\usepackage{graphicx}
\usepackage{float}
\usepackage{amsthm}
\usepackage{algorithm}
\usepackage{algpseudocode}
\usepackage{algorithmicx}
\usepackage{program}
\usepackage{caption}

\theoremstyle{definition}

\begin{document}

\maketitle

\section{Overview}

The core constructive task of the field 
of cryptography is that of creating cryptographic primitives (e.g. private-key encryption,
public-key encryption, message authentication codes,...) with provable security guarantees.
Where perfect (information theoretic) security is impossible, cryptographers must in practice necessarily rely on the assumption that some problem is hard to solve in nearly every 
case (a computational hardness assumption).  Primitives in these cases generally provide 
guarantees of the following informal form:
assuming that an enemy takes a long time to solve some problem $Y$, 
a system using $X$ is secure.

In general, the computational hardness assumptions used in the wild are derived 
from problems for which many have tried and failed to provide polynomial-time solutions; 
some of the most common examples include integer factorization, quadratic residuosity, 
the discrete log problem, and on.  At best, our reliance on these problems nests security 
in the hope that these specific, at times disparate problems will continue to elude 
researchers: by and large, none of these assumptions are supported by a well-founded 
general description of hardness as we would see in e.g. complexity theory.  As such, one of the so-called `holy grails' of cryptography is to 
instead vest the hopes for security in a single well-founded, well-known, and well-studied 
assumption, such as the assumption that $P \neq NP$.

While the reliable use of some NP-complete problem in tandem with the assumption that 
$P \neq NP$ has eluded cryptographers due to lack of results showing average-case hardness, 
one alternative which has been explored is reliance on assumptions that solving 
certain NP-hard optimization problems within some degree of accuracy 
is computationally difficult in specific instance 
classes.  In this work, we explore one such example of this effort, 
\cite{HidingCliques}, which attempts to construct cryptographic primitives by relying on the planted clique conjecture.  More specifically, we (1) present \cite{HidingCliques} in summary, 
(2) propose a simple cryptanalytic method for the one-way function primitive suggested in \cite{HidingCliques} 
using approximation algorithms, and (3) consider the feasibility of such cryptanalysis 
in the context of existing approximation algorithms for the maximum clique problem.
\tableofcontents

\section{Hiding Cliques for Cryptographic Security: Attempted Cryptography from the Planted Clique
Conjecture}

In ``Hiding Cliques for Cryptographic Security'' \cite{HidingCliques}, Juels and Peinado 
attempt to demonstrate a manner in which the problem of finding large cliques in random graphs 
may be used to construct cryptographic primitives.  Specifically, the authors note that it is 
conjectured that no probabilistic polynomial time algorithm exists which is able to 
find cliques of size $\geq (1+\epsilon) \log_2{n}$ for any $\epsilon > 0$ in random graphs 
with constant edge probability $\frac{1}{2}$ and attempt to develop cryptographic primitives
on the assumption that this conjecture is true.
\newline\newline
The authors begin with a general description of `hard' graph problems from the perspective 
of a standard variation of the clique problem in undirected graphs: given an undirected graph 
$G=(V,E)$ and an integer $k < \lvert V \rvert$, return a complete subgraph of $G$ a number 
of nodes as close as possible to $k$.  Following a discussion of the need for 
average-case hardness in cryptography, the authors then introduce the context of their own
contribution: where they acknowledge the previously stated conjecture regarding the ability to find `large' cliques 
in random graphs, the authors show that the problem of finding `large' cliques 
in random graphs with a constant number of artificially placed cliques (of size 
$1+\epsilon$, $\forall \epsilon, 1 \geq \epsilon > 0$) is just as hard as the truly random case
\footnote{What the authors rely on is simply a restatement of the planted 
clique conjecture--we discuss this in greater detail in a later section.}.  This contribution 
thereby admits a probabilistic method whereby hard instances 
of the clique problem may be generated and then used as the basis of cryptographic primitives. 

The authors do note that the actual security yielded by a naive application of this 
conjecture is weak.  In particular, the authors note that the largest natural clique 
size in a random graph with edge probability $\frac{1}{2}$ is $2 \log_2{n}$.  As such,
the brute-force approach needs only to iterate through all 
$\binom{n}{2 \log_2{n}} \in 2^{O(\log^2{n})}$ 
$2\log_2{n}$-node subsets of $V$ and therefore takes time pseudopolynomial in $n$; attacks 
relying on this brute-force approach may therefore be entirely practical.  The authors claim 
that this is a non-issue by virtue of the fact the complexity of the brute-force approach may 
be increased by increasing edge probability $p$ from $\frac{1}{2}$ (and thus the 
size of the largest clique); they justify this statement by further claiming that their main 
result holds for general edge probabilities.
\newline\newline
Following a discussion of related work largely restricted to the history of what we 
now know as the planted clique conjecture, the authors present a proof of their main result
which we summarize.  In the following proof sketch, the authors denote by $p$ the 
distribution $\mathcal{G}_{n,1/2}$ from the Erdos-Renyi random graph model (graphs 
consisting of $n$ nodes where each edge is present with probability $\frac{1}{2}$); they 
denote by $p'_k$ the distribution of graphs obtained by sampling from $p$ and uniformly 
selecting a subgraph of $k$ nodes to make complete.
\newline\newline
\noindent \textbf{Main theorem (T1) } For $k \leq (2-\delta) \log_2{n}$ and $2\geq \delta  > 0$,
if there exists an algorithm $A$ which finds cliques of size $(1+\epsilon)\log_2{n}$ with 
probability $\frac{1}{q(n)}$ (where $q(n)$ is a polynomial) in graphs drawn from 
$p'_k$, $A$ also finds cliques of size $(1+\epsilon)\log_2{n}$ with probability 
$\frac{1}{q'(n)}$ in graphs drawn from $p$ for some polynomial $q'(n)$.

\noindent \textbf{Proof of T1 } In their proof, given graph $G$, the authors denote by $C_k(G)$ the number of 
cliques of size $k$ in $G$.  For $G$ drawn from some distribution $D$, 
$C_k(G)$ is a random variable, and the authors refer to its expectation by $E_k$.

Because this result is not strictly crucial with respect to our analysis 
(our analysis will ultimately be indifferent to it), we simply provide a high-level sketch of the proof.  The first logical step in the proof is to show that, when $C_k(G\sim p'_k)$ is not too far 
from $E_k(G\sim p'_k)$, the probability of seeing $G$ when drawing inputs from $p'_k$ is negligibly 
far from that of seeing $G$ when drawing from $p$.  The next logical step is to prove that 
the variance of $C_k(G \sim p'_k)$ is satisfactorily small, thereby yielding that only a negligible 
fraction of graphs from the support of $p'_k$ will have $C_k(G\sim p'_k)$ far from 
$E_k(G\sim p'_k)$.  This allows them to conclude that there exists only a negligible 
fraction $\Delta$ of the support for which $p(G)$ and $p'_k(G)$ differ non-negligibly.  
Thus for the non-negligible $1-\Delta$-fraction of inputs which are close, we would expect a non-negligible difference in correctness when running $A$ on inputs drawn from this subset of the 
support according to $p$ (and thus a non-negligible difference in correctness overall).$\square$.
\newline\newline
\noindent The authors further claim (but do not prove) that T1 holds for general 
edge probabilities $p > \frac{1}{2}$.  Stated equivalently, they claim that the following 
also holds:  For $k \leq (2-\delta) \log_{\frac{1}{p}}{n}$ and $2\geq \delta  > 0$,
if there exists an algorithm $A$ which finds cliques of size $(1+\epsilon)\log_{\frac{1}{p}}{n}$ with probability $\frac{1}{q(n)}$ (where $q(n)$ is a polynomial) in graphs drawn from 
$p'_k$, $A$ also finds cliques of size $(1+\epsilon)\log_{\frac{1}{p}}{n}$ with probability 
$\frac{1}{q'(n)}$ in graphs drawn from $p$ for some polynomial $q'(n)$.
\newline\newline
\noindent After proving their primary result, the authors conclude the article by presenting 
a means of constructing cryptographic primitives using the assumption that there does 
not exist a probabilistic polynomial time algorithm able to find 
cliques of size $(1+\epsilon)\log_{\frac{1}{p}}{n}$ in graphs drawn from $\mathcal{G}_{n,p}$.
Of the primitives described, we focus our attention on their design of a secure  
\textbf{one-way function} \footnote{All of the constructions provided by the authors are 
provided in passing and without complete proof, but we focus on the OWF construction 
because it is arguably more versatile, 
able to be itself applied to obtain other primitives (e.g. symmetric-key encryption using a 
Feistel network).
(Furthermore, the other constructions they give are simply extensions of the OWF.)}.  
We reproduce the OWF described below:

\begin{algorithm}[H]
\caption{OWF from the Planted Clique Problem}\label{AMMDecode}
\begin{algorithmic}[1]
\Procedure{$f_{n,p,k}$}{$G=(V,E) \mid \lvert V \rvert = G \sim \mathcal{G}_{n,p}, K \subseteq V \mid \lvert K \rvert = k \in [\log_{\frac{1}{p}} n, 2\log_{\frac{1}{p}} n]$}
\State Connect all vertices in $K$, obtaining modified graph $G'$.
\State \textbf{return} $G'$
\EndProcedure
\end{algorithmic}
\end{algorithm}

\noindent (In the formal sense of a cryptographic one-way function, the first argument to 
$f$ would technically need to be an advice string $s$ sampling $G \sim G_{n,p}$ when $s$ is 
sampled uniformly.  (So it may be the case that $\lvert s \rvert \geq n$.).)  

\section{A Cryptanalytic Method using Approximation}

In this section, we (a) qualify how and under what specific assumptions the function $f$ 
proposed by Juels and Peinado is a cryptographic one-way function and (b) suggest a simple means of cryptanalysis of $f$ using nothing more than approximation algorithms.
\newline\newline
Formally, a cryptographic \textbf{one-way function} is a function $g$ which is (1) computable 
in polynomial time but (2) for which no probabilistic polynomial-time adversary $A$ 
(modeled as a randomized algorithm with source of randomness $r$) is able to 
compute a pseudo-inverse of $g$ except with negligible probability.  In other words,

\[ \forall \text{p.p.tm} A, \forall d > 0, Pr_r[g(A(r,g(x))) = g(x)] < n^{-d} \]

In the specific case of the candidate one-way function $f_{n,p,k}$ suggested by Juels and Peinado, 
we see that condition (1) is satisfied in that the operation of connecting $k$ vertices 
in an $n$-node graph may be trivially performed in polynomial time.  With respect to condition 
(2), consider the following modification to the computational variant of the 
\textit{planted clique conjecture} as given in \cite{PlantedCliqueSource}:

\begin{enumerate}
\item{Draw an Erdos-Renyi random graph $G$ from the distribution $\mathcal{G}_{n,p}$ for 
$p \geq \frac{1}{2}$.}
\item{Flip a coin to determine whether to modify $G$.  If heads, connect a random 
subset of vertices of size $k \in [\log_{\frac{1}{p}} n, 2\log_{\frac{1}{p}} n]$ in $G$; 
otherwise, leave $G$ as-is.}
\item{Return $G$.}
\end{enumerate}

\noindent The problem in this setting is, as usual, to find a clique in $G$ of size $k$ 
or output $\emptyset$ if none exists.

We claim now that $f_{n,p,k}$ satisfies condition (2) of a one-way function under the assumption 
that the computational planted clique problem is hard\footnote{as in it cannot be solved except 
with negligible probability by p.p.tm algorithms} in $\mathcal{G}_{n,p}$ and for 
$k$ in the range $k \in [\log_{\frac{1}{p}} n, 2\log_{\frac{1}{p}} n]$.  To show this, 
say that there exists a p.p.tm adversary $A$ which succeeds against $f_{n,p,k}$ with 
non-negligible probability.  In the planted clique setting, say that we flip heads in the 
second step.  Since the graph $G$ returned in the third step and the output of $f_{n,p,k}$ are both graphs drawn from $\mathcal{G}_{n,p}$ with a clique of size 
$k \in [\log_{\frac{1}{p}} n, 2\log_{\frac{1}{p}} n]$ embedded, $A$ will with high probability 
be able to recover a clique of size $k$ from the former because it is able to do so from the latter.

For $A$ which only ever returns cliques of size $k$ or the empty set, we thus see that $A$
succeeds in the planted clique setting with probability greater than or equal to half the probability it does so against $f_{n,p,k}$ as a one-way function, therefore non-negligibly by our 
choice of $A$.  By the contrapositive of this conclusion, we have that $f_{n,p,k}$ is a 
secure one-way function under the assumption that the computational planted clique problem is 
hard in $\mathcal{G}_{n,p}$ and for $k$ in the range $k \in [\log_{\frac{1}{p}} n, 2\log_{\frac{1}{p}} n]$.
\newline\newline
\noindent \textbf{Cryptanalysis through Approximation } We have so far established 
that the one-way function of Juels and Peinado is secure under the assumption that 
that the computational planted clique problem is 
hard in $\mathcal{G}_{n,p}$ and for $k$ in the range 
$k \in [\log_{\frac{1}{p}} n, 2\log_{\frac{1}{p}} n]$.  It follows directly, however,
that the inverse of this statement is also true: namely, if 
the computational planted clique problem can be solved in $\mathcal{G}_{n,p}$ and for 
$k$ in the range $k \in [\log_{\frac{1}{p}} n, 2\log_{\frac{1}{p}} n]$ with high probability, then 
$f_{n,p,k}$ is not secure.  We use this fact as a basis for cryptanalysis of $f_{n,p,k}$.

Say that we have an algorithm $B$ which is able to solve the computational planted clique problem 
in $\mathcal{G}_{n,p}$ for $k \in [\log_{\frac{1}{p}} n, 2\log_{\frac{1}{p}} n]$.  The output of 
$f_{n,p,k}(\cdot)$ is a graph drawn from $G_{n,p}$ with a clique of size $k$ embedded: since this 
corresponds to the heads case of the planted clique setting, $B(G)$ should will be able to 
output a clique $C$ of size $k$ with high probability.  This clique $C$ satisfies $f_{n,p,k}(G,B(G)=C) = f_{n,p,k}(\cdot)$, and so the existence of $B$ allows us to cryptanalytically invert $f_{n,p,k}$ 
in a direct manner.
\newline\newline
Extending the above analysis, we thus propose to cryptanalyze any practical implementation 
or use of $f_{n,p,k}$ as follows:

\begin{enumerate}
\item{Compile a finite, fixed set of algorithms $[A]$ approximating the maximum clique problem in graphs drawn from $\mathcal{G}_{n,p}$.}
\item{On obtaining an evaluation of $f_{n,p,k}(\cdot)=G'$, attempt to invert it by running 
$A(G)$ for all $A \in [A]$, returning the largest clique returned over all algorithms.}
\end{enumerate}

\noindent This method simply hedges cryptanalysis on the capabilities of approximation 
algorithms for the maximum clique problem in random graphs: if there exists an approximation 
algorithm which finds large cliques with high probability for some parameterization of 
$f_{n,p,k}$, we may succeed in inverting the function (or breaking any primitive which might 
be based on $f_{n,p,k}$) as we previously illustrated.    One might appreciate, however,
that cryptanalysis frees us from being tied to relying only on algorithms which 
require only polynomial time; a valid cryptanalytic attack may be conducted, for example, 
by a practical pseudo-polynomial algorithm which runs in parallel.  As such, 
$[A]$ may include anything from standard polynomial-time approximation algorithms 
to a clever variant of a brute-force approach.

In the rest of this report, we explore existing approximation algorithms 
to determine the extent to which such cryptanalysis is possible given our current capabilities.
We then use the information we uncover during this exploration to discuss the implications 
on methods of constructing cryptographic primitives with planted cliques in the manner of 
\cite{HidingCliques}.  

\section{Existing Approximation Algorithms}

We now shift our focus to survey existing strategies for searching for cliques in graphs.
We split this survey into two parts: (1) an exploration of methods constructed specifically 
for random graphs and (2) an exploration of the state-of-the-art with respect to approximation 
of the maximum clique problem.

\subsection{Methods for Planted Cliques in Random Graphs}

\subsubsection{The Greedy Approach}

One of the first formulations of the planted clique problem was given by Karp in 
\cite{PlantedCliqueGreedy}.  In his exposition, Karp gave a probabilistic analysis 
of a natural greedy approach to searching for cliques in random graphs.  This method 
in particular is interesting because of its simplicity and because more sophisticated 
methods achieve only comparable performance \cite{Metropolis}.  We reproduce the basic form of 
this algorithm below:

\begin{algorithm}[H]
\caption{Greedy Clique Search}\label{AMMDecode}
\begin{algorithmic}[1]
\Procedure{$f_{n,p,k}$}{$G=(V,E) \sim \mathcal{G}_{n,p}$}
\State Choose a random vertex $v \in V$.
\State Set $T$ to be the subgraph induced by $v$ and its neighborhood in $G$.
\State Set $C = \{ v \}$
\While{$T$ contains nodes not in $C$}
\State Choose a random vertex $u \in T$.
\State Add $u$ to $C$.
\State Set $T'$ to be the subgraph induced by $u$ and its neighborhood in $T$.
\State Set $T=T'$.
\EndWhile
\State \textbf{return} $C$
\EndProcedure
\end{algorithmic}
\end{algorithm}

\noindent \textit{Space and Time } It may be seen trivially that this algorithm 
requires polynomial time and space (polynomial with respect to $n$).  
There is one iteration per node in the 
set $C$ returned, of size at most $n$, and the operation to update $T$ may be 
performed naively in time $O(n)$ assuming $O(1)$ edge membership lookup, so 
the entire algorithm completes in time $O(n^2)$ (this is by no means tight).
Similarly, the algorithm needs only to store a constant number of copies of 
subgraphs of $G$, and so space space requireed is linear in the size of the original input.
\newline\newline
\noindent \textit{Approximation Performance } Karp gives a fairly direct 
probabilistic analysis which reveals that the given greedy algorithm will find a 
clique of size $(1+o(1)) \log_{\frac{1}{p}}{n}$ in expectation.  

Note first in the greedy algorithm that, at any iteration $C$ is a 
subset of $V$ and that $T$ is the subgraph of $G$ induced by $C$ and all nodes in 
$G$ which are adjacent to all nodes in $C$.  Note also that for each iteration 
in which the algorithm does not terminate $C$ grows by 1.  Since the algorithm 
only ever chooses nodes adjacent to all of $C$, $C$ is a clique of size $i$ during 
iteration $i$.

Draw a graph $G=(V,E)$ from $\mathcal{G}_{n,p}$.  Now fix an arbitrary subset 
$S_i \subseteq V$ of $V$, $\lvert S_i \rvert = i \leq n$; for our purposes, say 
$S_i$ is $C$ at the $i$th iteration of the greedy algorithm.  For all $u \in V \setminus S_i$, 
define an indicator random variable $N_u$ which is 1 if and only if $u$ is adjacent 
to all nodes in $S_i$.  Because edge probabilities are disjoint, we know 
$\mathbb{E}[N_u] = Pr[N_u] = p^i$ for all $u$; by linearity of expectation, we therefore 
may obtain 

\[ \mathbb{E}[\sum_{u \in V \setminus S_i} N_u] = (n-i) p^i \]

As a result, the earliest iteration $i$ such that the only nodes in $T$ 
are the nodes in $C$ (there are no more nodes adjacent to all of $C$) in expectation
satisfies $\mathbb{E}[\sum_{u \in V \setminus S_i} N_u] = (n-i) p^i = 0$, so 
$i = (1+o(1)) \log_2{\frac{1}{p}}{n}$.  During this iteration, $C$ is a clique of size 
$i$, and so, in expectation, the algorithm will return a clique of this size.  
\newline\newline
A further note which is perhaps more useful to our analysis than a general 
exploration of the planted clique problem is this: the variance in the 
number of nodes adjacent to all of $C$ in iteration $i$ obeys 
$Var[\sum_{u \in V \setminus S_i} N_u] = (n-i) (p^i) (1-p^i))$ because the indicators 
$N_u$ are not correlated and because they obey a Bernoulli distribution.  Thus, since 
$i$ brings $(n-i)p^i$ close to 0, it will likewise bring $(n-i)p^i(1-p^i)$ close to 0, 
meaning that there is little variance in the size of the clique returned by this 
greedy algorithm.

\subsubsection{The Metropolis Process}

In \cite{Metropolis}, Jerrum explores a method of applying a simplified form 
of simulated annealing called the Metropolis process to the planted clique problem.  
He ultimately finds that, though the method is more sophisticated, it requires 
super-polynomial time in order to find cliques larger than those found by greedy 
methods like the one discussed in the previous section.  
\newline\newline
Jerrum begins by describing what he calls the Metropolis process on cliques
in a graph $G=(V,E)$:

\begin{enumerate}
\item{At the current instant, let a clique $K$ be the `state' of our system.}
\item{Metropolis step:}
\begin{enumerate}
\item{Choose a node $v$ uniformly at random from $V$.}
\item{If $v \not\in K$ but $v$ is adjacent to all of $K$, add it to $K$.}
\item{If $v \in K$, remove $v$ from $K$ with probability $\lambda^{-1}$, where 
$\lambda$ is the parameter of the Metropolis process.}
\end{enumerate}
\end{enumerate}

\noindent The key difference between the Metropolis algorithm and a standard greedy 
algorithm is the $\lambda \geq 1$ temperature parameter: informally, it `tunes' the willingness 
of the algorithm to backtrack from local optima.  Jerrum also foreshadows properties 
of the method to be shown, noting the following properties of the process which 
make it seem like a plausible soluton to the planted clique problem:

\begin{itemize}
\item{The Metropolis process leads to an equillibrium distribution in which the 
probability of the current state being any particular clique $K$ is proportional to 
$\lambda^{\lvert K \rvert}$: in this way, the algorithm prefers larger cliques to 
smaller ones.}
\item{While increasing the $\lambda$ parameter to emphasize this difference 
does not come without a time trade-off, $\lambda$ doesn't need to be excessively large.}
\end{itemize}

\noindent The hope for the Metropolis process rests wholly in the desire for 
quick convergence; the author notes that this work dashes this hope, showing that 
the process takes times superpolynomial in $n$ to reach a state corresponding to 
a clique of size $(1+\epsilon) \log_2{n}$ \textit{regardless} of the choice of temperature 
$\lambda$.
\newline\newline
We now summarize the core analysis given by Jerrum.  Let $G=(V,E)$ be an undirected 
graph on $n$ nodes, and choose a temperature parameter $\lambda \geq 1$.  The Metropolis process 
on $G$ is discussed as a Markov chain $(G,\lambda)$ in which the state space 
$\Omega$ is the set of all cliques within $G$ and where there are transitions between states 
if and only if $\lvert K \Delta K' \rvert \leq 1$\footnote{Here, $\Delta$ means symmetric set difference.}.  The transition probabilities $K \Rightarrow K'$ in this chain are given by the following expression:

\begin{align*}
p(K,K') = \begin{cases}
\frac{1}{n} & \lvert K \Delta K' \rvert \leq 1 \land K \subseteq K' \\
\frac{1}{\lambda n} & \lvert K \Delta K' \rvert \leq 1 \land K' \subseteq K \\
1 - \sum_{K' \neq K} p(K,K') & K = K' \\
0 & \text{else}
\end{cases}
\end{align*}

Following this description of the process, Jerrum then delegates to elementary results 
in stochastic models to demonstrate that $(G,\lambda)$ has a unique stationary distribution 
$\pi(K) = \lambda^{\lvert K \rvert} \frac{1}{\sum_{K \in \Omega} \lambda^{\lvert K \rvert}}$.
\newline\newline
\noindent \textbf{Main result: A Lower Bound Argument } The main result of this work 
is stated formally as follows: for $G \sim \mathcal{G}_{n,p}$ and every $\lambda \geq 1$, there 
exists an initial state from which the expected time for the Metropolis process to reach a 
clique of size at least $(1+\epsilon) \log_2{n}$ exceeds $n^{\Omega(\log_2{n})}$.  

In order to prove this theorem, the author first introduces the notion of a \textit{m-gateway}, which is simply a clique from which there exists a path with non-zero transition probabilities to a clique of size $m$ in the Metropolis process.  The author then states and proves a lemma which establishes that for $k=(1+\frac{2}{3}\epsilon)\log_2{n}$, the proportion of $k$-cliques which are $m=(1+\epsilon)\log_2{n}$-gateways is less than $n^{-\Omega(\log_2{n})}$.  Logically, the purpose of this lemma within the proof is to show 
that $(1+\epsilon)$-cliques are generally inaccessible because $m$-gateway $k$-cliques are sparse
(and, for the chosen $m$, the algorithm must pass through some $k$-clique).

The proof of the main theorem then applies this lemma in a straightforward manner.
Given $(G,\lambda)$, and fixing $k$ and $m$ as in the lemma, we take $C$ as the 
set of $k$-cliques which are also $m$ gateways.  Using $C$, we then obtain a 
partition $(S,\bar{S})$ such that $S$ is the set of states that may be reached 
from the empty clique (initial state of the process) without passing through $C$.
The proof then uses lemma 1 and the fact that any $m$-clique must pass through a 
$k$-clique to show that the small size of $C$ constricts paths from $S$ to $\bar{S}$ 
according to the superpolynomial bound of $n^{-\Omega(\log_2{n})}$.
The authors then demonstrate formally that the expected time (number of iterations).
until the first entrance into $\bar{S}$ is itself therefore bounded from below by 
$n^{-\Omega(\log_2{n})}$, thus that there must necessarily exist an initial state which takes as long (and thus at least as long to reach an $m=(1+\epsilon) \log_2{n}$-clique located in $\bar{S}$).
\newline\newline
\noindent \textbf{Extensions: Larger Cliques and Varying Edge Densities } Jerrum 
also extends his main result to encompass cases corresponding to (a) the presence 
of large cliques and (b) the case where the edge probability is varied.  The 
primary argument follows the same pattern as the proof of the main theorem, and 
he is able to conclude in both cases that the superpolynomial bound holds.  More 
specifically, he shows the following:
\begin{itemize}
\item{For any $G \sim \mathcal{G}_{n,p}$ containing a clique of size 
$n^\beta$, $0 < \beta < \frac{1}{2}$ and every $\lambda \geq 1$, there exists an initial state from which the expected time for the Metropolis process takes time 
$n^{\Omega(\log_2{n})}$ to find a clique of size $(1+\epsilon)\log_2{n}$.
}
\item{For any $G \sim \mathcal{G}_{n,p}$, there exists an initial state from which the expected time for the Metropolis process takes time 
$n^{\Omega(\log_{\frac{1}{p}}{n})}$ to find a clique of size $(1+\epsilon)\log_{\frac{1}{p}}{n}$.}
\end{itemize} 

As we have seen thus far, there seems to be a `hard roof' with respect to the size 
of cliques found in strictly polynomial time by these methods hovering around 
$(1+o(1))\log_{1/p}{n}$.  While the variance admitted by the greedy approach does 
not offer much leeway, a key observation by Jerrum is that the theorems presented 
say something only about the absolute worst case of the Metropolis algorithm.  This 
point will become especially relevant in section 5.

\subsubsection{Finding Cliques using Spectral Methods}

In \cite{Spectral} authors Alon et. al. present a method of finding cliques 
of size $k > c n^{0.5}$ in graphs drawn from $\mathcal{G}_{n,1/2}$ 
for all fixed $c > 0$ using spectral methods.  Beginning with a 
discussion of intractibility results for the maximum clique problem in 
general graphs, the authors introduce motivation for studying large 
cliques planted in graphs from $\mathcal{G}_{n, 1/2}$ in that Jerrum 
\cite{Metropolis} and others conjecture the hardness of finding the maximum 
clique in graphs from $\mathcal{G}_{n,1/2}$ containing only natural cliques. 
\newline\newline
Let $G=(V,E)$ be an undirected  graph drawn from $\mathcal{G}_{n,1/2}$.  
Let $A$ be the adjacency matrix of $G$: $A_{u,v}=1$ if and only if $(u,v) \in E$.
Because $A$ is symmetric, it has real-valued eigenvalues 
$\lambda_1 \geq ... \geq \lambda_n$ and an associated orthonormal basis of 
eigenvectors $v_1,...,v_n$ (where $n$ is the number of nodes in $G$).
The authors initially present their method for $c$ fixed to 10
($k > 10 \sqrt{n}$); the 
key observation driving the algorithm is that, with high probability, a 
large portion of the hidden clique will be represented in the second 
eigenvector, $v_2$.  We reproduce the method below:

\begin{enumerate}
\item{Compute the the second eigenvector, $v_2$, of $A$ using any 
standard polynomial-time method.}
\item{Sort $V$ in decreasing order of absolute value in $v_2$.  Take
$W$ as the first $k$ nodes in this order.  Let $Q \subset V$ be the set 
of nodes in $V$ adjacent to at least $3k/4$ nodes in $V$.}
\item{Return $Q$.}
\end{enumerate}

That this algorithm returns in polynomial time is immediate.  There 
are many polynomial-time algorithms satisfactory for the first step, 
and the last two may be implemented as a simple sort/iterate procedure.  
In order to prove that the above algorithm returns a clique of size 
$k$, the authors prove some spectral properties of $G$; we provide an 
outline of the proof method and summarize the proofs given where appropriate.
The authors first prove the following:
\newline\newline
\noindent \textbf{Proposition 2.1 } Let $G \sim \mathcal{G}_{n,p}$ be a 
random graph; embed a clique of size $k = o(n)$.  Then almost surely 
the eigenvalues of the adjacency matrix $A$ satisfy that (i) 
$\lambda_1 \geq (\frac{1}{2} + o(1))n$ and (ii) $\lambda_i \leq (1+o(1))\sqrt{n}$
for all $i \geq 3$.

To prove (i), the authors use the common method of viewing $\lambda_1$ 
with respect to the Rayleigh quotient: 
$\lambda_1 = \max_{x\neq 0} \frac{x^TAx}{x^Tx}$.  Taking $x$ to be the 
vector of all 1s, we see thus that $\lambda_1$ is at least the average 
degree of nodes in $G$.  Since the degrees of $G$ may be described by 
a Binomial distribution determined by edge probability, we know that 
the average degree in $G$ is at least $(1/2 + o(1))n$ almost surely.

To prove (ii), the authors rely on the proven result that 
$\max_{i\geq 2} \lvert \lambda_i \rvert \leq \sqrt{m} + O(\sqrt{m}^{1/3} \log{m})$
for $G' \sim \mathcal{G}_{m,1/2}$ with high probability.  In order to 
bound the eigenvalues of $G$, the authors decompose $A$ as the union of 
two random graphs: $G_2 \sim \mathcal{G}_{k,1/2}$ and $G_1 = G - G_2$.  
Trivially, $G_1$ now obeys $\mathcal{G}_{n,1/2}$.  This then allows the authors
to decompose the adjacency matrix $A$ as a union of two adjacency matrices;
taking $u_1$ as the largest eigenvector of one and $u_2$ as the alrgest of 
the other, the largest Rayleigh quotient (over the subspace of vectors 
orthogonal to both $u_1$ and $u_2$) will with high probability be 
$\leq (1+o(1))\sqrt{n}$ by the afforementioned existing result.
\newline\newline
The rest of the proof rests on the crucial point to be made by the analysis 
that the eigenvector $v_2$ of $A$ has the majority of its weight 
on the large clique in $G$.  To show this point, the authors 
define a vector $z \in \mathbb{R}^{n}$ such that $z_i = n-k$ if 
$i \leq k$ and $-k$ otherwise.  Towards showing that the weight of 
$v_2$ rests in the clique of size $k$, the next step in the proof is to 
show the following proposition to be true:
\newline\newline
\noindent \textbf{Proposition 2.3 } There exists a small vector $\delta$
(specifically having $l2$-norm less than $\frac{1}{60}$th of that of $z$)
such that $z-\delta$ is collinear with $v_2$.

The proof of this proposition makes use of the following lemma: with 
high probability,

\[ \lvert \lvert (A - (k/2)I)z \rvert \rvert^2 \leq (1/4 + o(1))n^3k \]

\noindent While we will not reproduce the proof of this lemma in 
great detail, it follows fairly directly.  In order to prove it, one 
must simply decompose the coordinates of $(A - (k/2)I)z$ as 
random variables; the claim of the lemma then follows by applying 
standard Binomial estimates.

The truth of the proposition follows from this lemma using 
standard spectral decomposition.  Decompose $z$ as a linear 
combination of the vectors of the orthonormal eigenbasis of $A$: 
$z=c_1v_1+...+c_nv_n$.  Likewise, take $\delta = c_1v_1+c_3v_3+...+c_nv_n$.
Note that then $(A-(k/2)I)z = \sum_{i=1}^{n} c_i(\lambda_i - (k/2))v_i$, so 

\begin{align*}
\lvert \lvert (A-(k/2)I)z \rvert \rvert&=\sum_{i=1}^{n} c_i^2(\lambda_i - (k/2))^2\\
&\geq (1+o(1))(\frac{k}{2} - \sqrt{n})^2 \sum_{i=1}^{n} c_i^2
\shortintertext{(Prop 2.1) (R1)}
\end{align*}

\noindent Using this result, the authors then show that $\lvert \lvert \delta \rvert \rvert^2$
is less than $\frac{1}{60}\lvert\lvert z \rvert\rvert^2$:

\begin{align*}
\lvert \lvert \delta \rvert \rvert^2 &= \sum_{i \neq 2}^{n} c_i^2 \\
& \leq (1+o(1))\frac{n^3k}{(k-2\sqrt{n})^2} \\
\shortintertext{(Rearranging R1; applying lemma)} \\
& < \frac{1}{60}kn(n-k) = \frac{1}{60}\lvert\lvert z \rvert\rvert^2
\end{align*}

\noindent Within the larger scope of the analysis of this algorithm, 
proposition 2.3 uses proposition 2.1 to obtain that there exists a 
vector $z-\delta=c_2v_2$ collinear with $v_2$ which is `heavy' 
in $k$ dimensions. The authors conclude the proof of correctnesss 
by establishing the following two claims:

\begin{enumerate}
\item{At least $\frac{5}{6}$ of the $k$ largest coordinates of 
$v_2$ correspond to vertices of the clique.  The authors argue that, 
because $\lvert \lvert \delta \rvert \rvert^2 \leq (1/60)\lvert \lvert z^2 \rvert \rvert$, there are at most $k/6$ coordinates of $\delta$ which are greater 
than $n/3$ in absolute value.  By the structure of $z$, then, the authors 
are able to conclude that at least $k-d_1$ of the first $k$ coordinates 
have value greater than $n/2$ and at least $n-k-d_2$ of the last $(n-k)$ 
coordinates have value at most $n/2$, where $d_1 + d_2 \leq k/6$.  The 
authors claim that this then implies the first statement of this item. 
\footnote{We note that this indeed follows if the first $k$ coordinates 
correspond to the elements in the $k$ clique, which the authors haven't 
explicitly established.  The argument certainly holds if we arbitrate 
that the first $k$ nodes in the graph are the planted clique, and it 
should hold under permutation of vertices; the authors do not discuss 
this, but it is only a minor point.}}
\item{Every vertex outside of the clique is almost surely adjacent to 
less than $3k/4$ vertices of $W$.  The authors establish this through 
a probabilistic argument: owing to edge probability $\frac{1}{2}$, 
vertices outside of the clique are adjacent to at most $(1+o(1))k/2$ 
vertices of the clique.  Since $W$ also contains at most $1/6$ nodes 
not in the clique, we may bound the maximum number of neighbors in
$W$ of any non-clique node by $(1+o(1))k/2 + k/6 < \frac{3}{4}k$.}
\end{enumerate}

\noindent Together, these two claims conclude the proof of correctness 
of the algorithm.  This guarantees (with high probability) by (2) that the 
algorithm chooses only vertices of the clique and by (1) that the algorithm
takes all vertices of the clique.
\newline\newline
The final section of \cite{Spectral} gives a method of extending the 
original algorithm in order to find cliques of size $c \sqrt{n}$ for 
fixed $c$.  The main observations leading to this improvement are the 
following:

\begin{itemize}
\item{By taking the subgraph induced by the neighborhood of a 
subset of vertices $S$ of size $s$, 
$G[S]$, we reduce the size of the graph to almost certainly contain only 
$(1+o(1))n/2^s$ nodes.}
\item{When the subgraph is induced by a subset of nodes $S$ is 
also a subset of the largest clique in the graph, the ratio of the 
clique size to the number of the nodes in the graph increases.}
\end{itemize}

\noindent Noting these, the authors conclude by presenting the following 
polynomial-time method which extends their original algorithm:

\begin{enumerate}
\item{Calculate $s = 2\log_2{10/c} + 2$.}
\item{For all subsets $S \subset V$ of size $s$}
\begin{enumerate}
\item{Calculate $G'$ as the subgraph induced by $S$ and nodes in the intersection 
of neighborhoods of all nodes in $S$.}
\item{Run the original algorithm on $G'$, obtaining a set $Q_S$.}
\item{If $Q_S \cup S$ is of size $k$, return $Q_S \cup S$.}
\end{enumerate}
\item{Return $\emptyset$.}
\end{enumerate}

\subsubsection{A Probabilistic Combinatorial Method}

Since the publication of \cite{Spectral}, other authors have found alternative
approaches which allow for the recovery of cliques of size $c\sqrt{n}$ in 
random graphs.  One example is \cite{SDP}, in which Feige describes another 
method achieving the same bound as \cite{Spectral} using semi-definite programming.

Another example which we will explore in greater detail in this section is 
a method given by  Dekel et. al. in \cite{Combinatoric}.  This method is of particular interest to us because (a) it achieves performance identical 
to that achieved in \cite{Spectral}, (b) it relies on a simple 
randomized combinatorial strategy, and (c) it is easily extended to apply 
to random graphs with edge probability other than $\frac{1}{2}$.

At a high level, for a graph $G \sim \mathcal{G}_{n,p}$ with a hidden 
clique of size $k$, the algorithm proposed by Dekel et. al. is parameterized
by $0 < \alpha < 1$ and $\beta > 0$ and proceeds in three stages:

\begin{enumerate}
\item{Iteratively find subgraphs of the input graph $G$:
$G=G_0 \subset G_1 \subset ...$.  In particular, given $G_i$, 
$G_{i+1}$ is defined as follows: pick a random subset of vertices 
$S_i$ from $G_i$ which contains each vertex with probability $\alpha$.  
Define $\bar{V}_i$ to be the set of vertices in $G_i$ that are not 
in $S_i$ but have at least $\frac{1}{2}\lvert S_i \rvert + 
\beta\frac{\sqrt{\lvert S_i \rvert}}{2}$ neighbors in $S_i$.  
Intuitively, this process grows the relative size of the 
hidden clique with each iteration; in general, this process needs 
to be repeated a logarithmic number $t$ times.}
\item{Search for $\bar{K}$, the subset of the hidden clique 
contained in $G_t$.  This is done by computing an estimate 
of the size of $\bar{K}$, $k_t$, and taking $\bar{K}$ as 
the set of vertices in $G_t$ with degree at least 
$\frac{1}{2}\lvert V(G_t) \rvert + \frac{1}{4}k_t$.}
\item{Find the rest of the nodes of $K$ using $\bar{K}$.
This is done by taking $K'$ as the set containing $\bar{K}$
and all of its common neighbors.  $K^*$ is taken as the 
set of the $k$ largest-degree vertices in the subgraph of $G$ 
induced by $K'$.  $K^*$ is the set returned by the algorithm 
as the hidden clique.}
\end{enumerate}

\noindent The main result that the authors prove about this 
algorithm is the following:
\newline\newline
\noindent \textbf{Theorem } For $c > c_0$, there exist parameters
$\alpha$ and $\beta$ such that, for $G\sim\mathcal{G}_{n,1/2}$ with 
a hidden clique of size $c\sqrt{n}$, the probability that 
$K^*(G,\alpha,\beta)$ is the hidden clique is at least 
$1-e^{-\Theta(n^{\epsilon_0})}$ where $\epsilon_0$ is a function 
of $c$.

\noindent We provide a sketch of the proof given by the authors in 
\cite{Combinatoric}.  

With respect to phase (1), the 
authors consider an iteration $i$ of the subgraph-finding procedure
to be successful if the number of nodes in $G_i$ and the 
number of nodes of the hidden clique in $G_i$ are each respectively 
close to their expected value; the authors show that this is 
the case with high probability.  Specifically, the authors 
show that, in every iteration, the graph $G_i$ has the same 
distribution as random graphs $\mathcal{G}_{\bar{n}_i,1/2}$ 
with cliques of size $\bar{k}_i$ embedded (and where 
$\bar{n}_i$ and $\bar{k}_i$ are the respective expected values).
This allows the authors to conclude that the per-iteration 
failure probability is small enough to take a union bound 
over all $t$ iterations and obtain a high success probability.

With respect to phase (2), the authors show that the set 
$\bar{K}$ is with high probability a subset of the hidden clique.
They do this by proving two key lemmas.  The first of these 
lemmas states that, for random graphs with embedded cliques of
a certain size, there exist natural numbers $D_1$ and $D_2$ 
such that all non-clique nodes have degree less than or equal 
to $D_1$ and all clique nodes have degree greater than or 
equal $D_2$.  The second of these lemmas uses the previous 
to state that, if phase (1) completed without failure 
(which occurs with high probability due to their proof 
of the properties of phase (1)), the nodes of the 
hidden clique are distinguishable via $D_1$ and $D_2$ and thus 
$\bar{K}$ is with high probability a subset of the original hidden clique.

With respect to phase (3), the authors give a more 
general result that the entire hidden clique may be revealed 
given partial information in the form of a subset of the hidden 
clique.  In particular, the authors show that the entire hidden 
clique may revealed if the size of the hidden clique $k$ and the 
size of the partial clique revealed $s$ has the following properties:

\begin{itemize}
\item{$k \in O(\log{n}\log{\log{n}})$ and $s \geq (1 + \epsilon) \log{n}$}
\item{or $k \in \omega(\log{n}\log{\log{n}})$ and $s \geq \log{n} + 1$.}
\end{itemize}

\noindent by the proof of the properties of phase (2), one of the 
above is the case with high probability, and so $K^*$ obtained
from $\bar{K}$ is the hidden clique.
\newline\newline
Following their proof of the basic algorithm, the authors 
give multiple extensions of their algorithm, two of which are useful 
for our purposes.
\newline\newline
\noindent \textbf{Reducing $c_0$ } The authors note 
that the basic form of their algorithm works for 
$c \geq c_0$ for $c_0 \geq 1.65$\footnote{The authors also 
introduce a more complicated method of reducing $c_0$ to about 
1.22, but the performance gain is marginal in the face of 
the method of Alon in \cite{Spectral}}.  In their conclusion, the authors 
note that the same technique used by Alon et. al. in \cite{Spectral}
to reduce their $c_0$ of 10 applies: simply iterate through 
subgraphs $G[N^*(S)]$ induced by the common neighborhoods of small 
subsets of nodes and run the original algorithm on each 
subgraph.
\newline\newline
\noindent \textbf{Alternative Edge Probabilities } The authors 
further provide a generalization of their original algorithm 
which is able to in polynomial time find cliques of size $c\sqrt{n}$ 
for all $c > c_0$ and for graphs with edge probabilities
$p \neq \frac{1}{2}$.  We summarize the differences in this more 
general form of the algorithm below:

\begin{enumerate}
\item{Run phase 1 as before, except now take 
$\bar{V}_i$ as the set of vertices with at least 
$p\lvert S_i \rvert + \beta \sqrt{p(1-p)\lvert S_i \rvert}$ 
neighbors in $S_i$.}
\item{Define 
$\rho=(1-\alpha)\bar{\Phi}(\beta-c\sqrt{\alpha}\frac{1-p}{\sqrt{p(1-p)}})$.  After phase 1 has run for $t$ iterations, let 
$\bar{K}$ contain all of the vertices in $G_t$ with degree 
at least $p\lvert V(G_t)\rvert + \frac{1}{2}(1+p)\rho^tk$.}
\item{Phase 3 is as before, but instead let $K'$ contain $\bar{K}$ 
and all vertices in $G$ having $\frac{1}{2}(1+p)\lvert \bar{K} \rvert$
neighbors in $\bar{K}$.  Let $K^*$ now be the set of vertices 
in $G$ having at least $\frac{1}{2}(p+1)k$ neighbors in $K'$.}
\end{enumerate}

\noindent Note that above $\bar{\Phi}()$ refers to the complementary Gaussian cumulative distribution function. The proof of this generalized algorithm is the same as before, except the variable $p$ is kept as a parameter rather than manipulated as a constant $\frac{1}{2}$.  The authors do not offer a 
formal analysis of algorithm runtime, but their stated running time is 
$O(n^2)$ for all given algorithms.  (This excludes the case for the 
algorithm boosted using the subset enumeration technique of Alon from 
\cite{Spectral}.)

\subsubsection{Recent work: Sum-of-squares Optimization}

Another way to solve the planted clique problem is by relaxing the clique problem to sum-of-squares (SOS) optimization. A Sum-of-squares optimization program has the following form: \newline
\begin{align*}
max_{u \in \mathbb{R}^n} c^Tu & \\
& \text{s.t. } \forall k = 1, ..., J & a_{k,0}(x) + a_{k,1}(x)u_1 + ... + a_{k,n}(x)u_n \in \text{SOS forms}\\
\end{align*}
Above, ``SOS forms'' refers to the set of polynomials of the form: 
\begin{align*}
	f(x) = \sum_{i=i}^{k} g_i(x)^2 
\end{align*}
\noindent where $g_i$ are polynomials whose non-zero terms are all of the same degree. 
SOS programs are generally non-convex, and the standard method of solving them relies on 
convex relaxations (e.g. through relaxing constraints to semi-definite constraints on 
matrices). More specifically, there exists a hierarchy of convex relaxations called the
SOS hierarchy which is indexed by a natural number parameter $d$: as $d$ increases, 
solving power increases at the expense of increased computational cost. 

A recent paper by Barak et. al. looks at the limitations of SOS optimization when 
used to address the planted clique problem \cite{SOS2016}.  The only method of solving the 
SOS relaxation of the planted clique problem at the $d$th level of the hierarchy 
cited in this work (and the related lower bounds papers which they cite) requires time 
$n^{O(d)}$ and involves expressing the program as a SDP in $n^{O(d)}$ variables. The authors claimed to have derived a somewhat tight lower bound on the size of cliques able to be found in 
graphs drawn from $\mathcal{G}_{n,1/2}$ for a given degree $d$. The bound was shown to be $n^{\frac{1}{2}-c(\frac{d}{\log n})^{\frac{1}{2}}}$ and thus for any degree $d = o(\log n)$ is $n^{\frac{1}{2} - o(1)}$.  

To achieve this bound, the authors apply a computational variant of standard Bayesian probability. Barak et al. explain that another way to view SOS optimization is as a way to get a set of internally consistent computational probabilities. The probabilities of interest in the planted clique problem refer to probabilities of nodes being in the clique given that another node is or is not in a clique or the fact that nodes with larger degrees are more likely to be in the clique. These probabilities can be seen as giving rise to a pseudo-distribution which can then be used to prove properties about the solution space of an 
SOS program relaxed to the $d$th level of the hierarchy.  Because 
\cite{SOS2016} provides only a lower bound on a specific approach, we do not explore the proof 
of the authors in detail; however, we note that the authors achieve their bound by illustrating 
that the ability to distinguish large cliques using these pseudo-distributions degenerates 
as the clique number $k$ is less than $n^{1/2-\delta}$.

\subsection{Methods for General Graphs}

The previous sub-section concerned itself with exploring methods catering specifically to random graphs 
drawn according to the Erdos-Renyi model.  Another obvious family of algorithms relevant to the OWF $f_{n,k,p}$
is the family of algorithms concerned with finding approximately the largest clique in a \textit{general} graph 
$G$.  Because the maximum clique problem is known to be hard to approximate with a factor of $O(n^{1-\delta})$ unless
$P = NP$ \cite{HardToApprox}, we restrict our exploration of this class of algorithms to the single best-known 
approximation algorithm for the maximum clique problem\cite{BestKnown}.

\subsubsection{Feige's Algorithm: The State of the Art}

The final algorithm is an approximation of maximum clique by Feige that achieves an approximation factor of $O(n(\log{\log{n}})^2/\log^3{n})$ in general graphs \cite{FeigeApprox}. To get this ratio, Feige first proposes a new algorithm that can generally find cliques of at least size $t\log_{3k}(\frac{n}{t})-3$ given that the graph has a clique of size $\frac{n}{k}$. When $t = \omega(\frac{\log n}{\log \log n})$ the proposed algorithm can find cliques of size $(\frac{\log n}{\log{log n}})^2$ in polynomial time given the graph has a clique with size at least $\frac{n}{(\log{n})^b}$ for some arbitrary $b$. The author then uses this algorithm in conjunction with Boppana and Halldorsson's existing algorithm \cite{Boppana1992} to achieve the approximation ratio.

We now describe the Feige's presentation of his algorithm.  Let $G(V,E)$ be a graph with $n$ vertices with a clique of size $\frac{n}{k}$. Let $t << \frac{n}{k}$ be a parameter that inversely determines the final size of the clique and the runtime.  Before enumerating the steps of the algorithm, Feige 
gives the following definition and lemma which we reproduce and summarize here.
\newline\newline
\noindent \textbf{Definition 1 } Any vertex induced subgraph $S$ of $G$ that does not have cliques of size at least $\frac{\mid S\mid }{2k}$ is \textit{poor}.
\newline\newline
\noindent \textbf{Lemma 1 } If G(V,E) has a clique of size $\frac{n}{k}$. Then the vertex induced subgraph of G with all poor subgraphs removed, G'(V', E'), will contain a clique of at least size $\frac{\mid V'\mid }{k}$ and $\mid V'\mid  \geq \frac{n}{2k}$
\newline\newline
The general intuition behind the proof of the lemma above comes from the observations that (1) the union of poor subgraphs results is a poor subgraph and
that (2) removing a poor subgraph from $G$ will increase the density of the maximum clique in G'.  The claim of the lemma then follows directly.
\newline\newline
\cite{FeigeApprox} then begins to explain the algorithm at a very high level.  Intuitively, the algorithm proceeds in a sequence of phases, 
at the end of which one of the following is true:
\begin{enumerate}
\item{A clique of size $tlog_{3k}(\frac{\mid V'\mid }{k})$ if found and the program terminates.}
\item{A poor subgraph is found and removed from $G'$ (or initially $G$). Start new phase.}
\end{enumerate}
Using lemma 1, we know that, at the start of each phase (after the subgraph is removed), the size of the clique in $G'$ will be at least $\frac{\mid V'\mid }{k}$. Lemma 1 also tells us that $\mid V'\mid $ will be $\geq \frac{n}{2k}$, so this algorithm must terminate and give us a final clique of at least size $t \log_{3k}(\frac{n}{12k^2t}) > t\log_{3k}(\frac{n}{t} - 3)$.
Each phase has several iterations that each take as input a subgraph of $G'$, $G"(V", E")$, (or initially just $G'$) and a set of vertices $C \in V' \ V"$ that create a clique (or initially an empty set). Each iteration works as follows:
\begin{enumerate}
\item{If $\mid V"\mid  < 6kt$, return $C$ as clique}
\item{Partition $V"$ into disjoint parts of size $2kt$ vertices.}
\item{For each part $P_i$}
	\begin{enumerate}
		\item{For each subset of vertices $S_{ij}$ of $P_i$ with cardinality t}
			\begin{enumerate}
				\item{Let $N(S_{ij})$ be the set of vertices $\in V" \ S_{i,j}$ that are connected in G" to every vertex in $S_{ij}$}
				\item{If the subgraph of $G"$ induced on $S_{ij}$ is a clique and $\mid N(S_{ij}) \geq \frac{\mid V"\mid }{2k - t}$, then let $C = C \union S_{ij}$ and let $G"$ be the subgraph induced on $N(S_{ij})$. Return and start the next iteration.} 
			\end{enumerate}
			\item{Else return V" as as a poor subgraph.}
	\end{enumerate}
\end{enumerate}
Feige then analyzes the run-time. Each iteration will remove at least $t$ vertices from $V'$, so there are $\mid V'\mid /t$ maximum iterations. In each iteration, there are $\binom{2kt}{t}$ subsets to go through, so the whole phase will be polynomial if $\binom{2kt}{t}$ is polynomial in $n$. So, given the choice of $t$ mentioned earlier, it will be polynomial in run-time. 

Feige then uses proof by contradiction to shows that if a phase says $V"$ is poor, that it is actually poor (i.e. does not contain a clique of size $\frac{\mid V"\mid }{2k}$). He assumes there is a clique of size $\frac{\mid V"\mid }{2k}$ and applies the pigeon-hole principle to say that one of the parts $P_i$ must contain at least $t$ vertices from this clique and thus $V"$ would not be returned as a poor subgraph.  

The author then proves that the phase outputs $C$, then $C$ must be a clique in $G'$ and have at least size $t \cdot log_{3k}(\frac{n'}{6kt})$. $C$ must be a clique in $G'$ because everything unioned with $C$ was a clique and based off our modification of $G'$, all later nodes that are being considered to be added to $C$ will have been neighbors to everything in $C$. The size comes from the fact that each iteration adds $t$ vertices to $C$ and the number of iterations to get $\mid V'\mid $ to $6kt$ is at least $log_{3k}(\frac{n'}{6kt})$.

Feige then moves on to the final, complete presentation of his approximation algorithm which makes use of three cases. If the size of the maximum clique is $ < \frac{n}{(\log n)^3}$, output a single vertex to obtain an approximation ratio of $O(\frac{n}{(\log n)^3})$. If the size of the maximum clique is larger than $\frac{n}{(\log n)^3}$, use the previous algorithm to obtain an $O(\frac{n(\log \log n)^2}{(\log n)^3})$ approximation ratio in polynomial time (the size of maximum clique is $O(\frac{n}{log n})$). If the size of the maximum clique is larger than $\frac{n}{log n}$, we need to modify the previous algorithm by using Boppana and Halldorsson's algorithm \cite{Boppana1992} in order to find cliques larger than $(\frac{\log n}{log log n})^2$. Their algorithm relies on the fact that any graph with n = $\binom{s + r - 2}{s - 1}$ will either have an independent set of size $r$ or a clique of size $s$ and then gives the algorithm to find either the independent set or the clique \cite{Boppana1992}. For the modified algorithm, if the Boppana and Halldorsson's algorithm returns a clique, then use that as the output, otherwise if it returns an independent set, then use the independent set to discover a poor subgraph and remove it. This leads to the following proposition: 
\newline\newline
\noindent \textbf{Proposition 1 } Given a graph with a clique larger than ${2n \log \log n}{log n}$, the above algorithm will find a clique of at least size $\frac{(\log n)^3}{6 \log \log n}$
\newline\newline
From the proposition, we get that using the above will given an $O(n(\frac{\log \log n}{\log n})^3)$ approximation ratio. Note that this approximation ratio, $O(n(\frac{\log \log n}{\log n})^3)$, is a factor $\omega(\log \log n)$, off from the $O(n\frac{(\log \log n)^2}{(\log n)^3})$ goal. So to save another $\omega(\log \log n)$, Feige adapts another technique from Halldorsson \cite{Hallorsson:1993}. Instead of checking that $\lvert N(S_{ij}) \rvert \geq \frac{\mid V"\mid }{2k - t}$ in the detailed algorithm, we instead check $\lvert N(S_{ij})\rvert  > n_{test} - t$ where $n_{test}$ is the largest value satisfying $n_{test} < (\frac{\log n_{test}}{2 \log \log n_{test}})(\frac{\mid V"\mid }{2k})$. In the case that $\frac{\mid V"\mid }{2k} - t \leq \lvert N(S_{ij})\rvert  \leq n_{test} - t$, run the Boppana and Halldorsson algorithm on the subgraph of $S_{ij} \cup N(S_{ij})$. If it returns a clique of at least size $\frac{log n_{test}}{6 \log \log n_{test}}$, then add it to C and return. Feige then states that the analysis of this modified algorithm is similar to that of the originally proposed algorithm.  We don't explore this in greater detail because this does not bring the approximation ratio down to an interesting point with regard to the planted clique problem 
(this will become clear when we attempt to apply it in cryptanalysis). 

\section{Applying the Cryptanalytic Method}

In this section, we take $[A]$ to be the set of algorithms we have explored 
thus far and determine the circumstances under which each would allow us to 
compute the pseudo-inverse of  the one-way function $f_{n,p,k}$ suggested in 
\cite{HidingCliques} with non-negligible probability (thus `breaking' the primitive).
We then use this analysis to tabulate our conclusions in summary and give 
recommendations for secure function parameters given a desired level of 
security $\lambda$ and our current capabilities with respect to the planted clique 
problem; our analysis also allows us to suggest future work as regards 
developing the primitive further to address practical concerns and potential 
vulnerabilities.

\subsection{Limits of Analysis given Existing Approximation Algorithms}

For each algorithm explored, we distill the circumstances under which the 
algorithm is able to compute the pseudo-inverse of $f_{n,p,k}$.  For convenience, 
we remind the reader that $n$ is the number of nodes in the graph returned, 
$p$ is the edge probability within the graph generated, and 
$k = [\log_{1/p}{n}, 2\log_{1/p}{n}]$ is the size of the clique planted in the graph returned.

Our consideration of the security of $f_{n,p,k}$ will be done with respect to a 
parameter $\lambda$.  $\lambda$ is the desired security parameter of the system, 
and it is defined as follows: any non-parallelized attack against $f_{n,p,k}$ must
require at least $2^\lambda$ steps.  In its essence, $\lambda$ is a measure 
of how hard $f_{n,p,k}$ is to invert for any adversary.
\newline\newline
\noindent \textbf{The Brute-Force Approach } The original paper suggesting the
one-way function $f_{n,p,k}$ admitted that the brute-force search algorithm,
though not polynomial in runtime, is still practical.  The basic approach 
of an attack leveraging a brute-force search would involve simply enumerating
all subsets of nodes of size $2\log_{1/p}{n}$ and checking whether the subset 
is completely connected.  In order to satisfy the security parameter 
$\lambda$, we need this process to take at least $2^\lambda$ steps.
Thus, one solution is as the authors describe: increase $p$ until 
$k$ has size $n^d$ for some constant $d$ such that 
$\binom{n}{n^d} > 2^\lambda$; in particular, if we consider only brute-force attacks,
we suggest increasing 
$p$ to be $2\log_2{\lambda}/\lambda$ and setting $\lambda=n$ so that the embedded clique size is 
around $\lambda/2$.  Another solution is to keep $p=\frac{1}{2}$ 
and increase $n$ until $\binom{n}{\log_2{n}} \geq 2^{\log^2{n}-\log{\log{n}}} 
\geq 2^\lambda$.  In this case, it would be sufficient to set $n = 2^{\Omega(\sqrt{\lambda})}$.
This, however, is less desirable than the first solution, as it requires the generated graph to 
be exponential in the square root of the security parameter.
\newline\newline
\noindent \textbf{The Greedy Approach in Random Graphs } In section 4.1.1, 
we saw an analysis of a simple randomized greedy approach to searching 
for large cliques in random graphs drawn from $\mathcal{G}_{n,p}$.  This 
analysis showed that the algorithm is able to in time $O(n^2)$ (where $n$ is 
the number of nodes) find cliques of size $(1+o(1)) \log_{1/p} n$ in expectation 
and that the variance of the returned clique size is very small.  We thus see 
that, for general $n$ and $p$, $f_{n,p,k}$ is vulnerable to the greedy approach
only when $k=(1+\epsilon)\log_{1/p}{n}$ for $\epsilon$ very close to 0.
\newline\newline
\noindent \textbf{The Metropolis Process in Random Graphs} In section 4.1.2,
we saw a randomized method based on simulated annealing for searching for 
large cliques in random graphs drawn from $\mathcal{G}_{n,p}$.  Where $\phi$ 
is the temperature parameter, the result we saw was that, for any temperature 
parameter $\phi$ and $\epsilon > 0$, there exists an initial state of the algorithm for which 
the algorithm requires $n^{\Omega(log_{1/p}{n})}$ time to find a clique of size
$(1+\epsilon) \log_{1/p}{n}$.  This implies that, for the worst possible initial 
configuration, the algorithm can only find cliques of size less than or equal to 
$\log_{1/p}{n}$.  In this worst-case setting, we thus see that for general 
$n$ and $p$, $f_{n,p,k}$ is vulnerable to the Metropolis process only when 
$k=(1+\epsilon)\log_{1/p}{n}$ for $\epsilon$ very close to 0.  As noted in the next 
paragraph, however, the results of the original paper give no indication of time 
taken in the average case.

There are two caveats which will be important later in our analysis: namely, the authors prove that 
(1) \textbf{there exists} an initial state for which it is hard to find large cliques
and (2) there is no result on how this method performs in the average case or 
with partial information.

We shelf the issue of lack of average-case results for this method and for now 
consider the security of $f_{n,p,k}$ against Metropolis as-is and in the worst case as 
regards efficiency.  In order to satisfy a security level of $\lambda$, we need the figure of 
$n^{\Omega(log_{1/p}{n})}$ to translate to a load of at least $2^\lambda$ when searching for 
cliques in the output of $f_{n,p,k}$.  We see by the following:

\begin{align*}
&n^{\log_{1/p}{n}} &\geq 2^{\lambda} \\
\Rightarrow& 2^{\log{n}\log_{1/p}{n}} &\geq 2^{\lambda} \\
\Rightarrow& \log{n}\log_{1/p}{n} &\geq \lambda \\
\Rightarrow& \log^2{n} &\geq \log_{1/p}\lambda \\
\Rightarrow& n &\geq 2^{\sqrt{\log_{1/p}\lambda}}
\end{align*}  

\noindent that we may force this to be the case by setting $n = 2^{\Omega(\sqrt{\log{1/p}\lambda})}$.
\newline\newline
\noindent \textbf{Spectral Methods in Random Graphs } In section 4.1.3,
we saw a means of searching for large cliques embedded in graphs 
drawn from $\mathcal{G}_{n,1/2}$ using spectral techniques.  
The analysis of the method showed that 
it may be manipulated in order to find large cliques of size 
$c\sqrt{n}$ for any fixed $c$ in polynomial time.  Since their method is also
catered specifically for edge probability $\frac{1}{2}$, we are thus able to conclude 
for this method in particular that $f_{n,p,k}$ is vulnerable to this spectral 
approach only in the case that $p=\frac{1}{2}$ and $k \in \Theta{\sqrt{n}}$.

Although $k \in [\log_2{n}, 2\log_2{n}]$ for $p=\frac{1}{2}$ and therefore 
this vulnerability will not hold asymptotically for $n$ large enough, it is important
to note that the method of Alon in \cite{Spectral} works even for cliques of size 
$c \sqrt{n}$, $c < 1$, using partial information found through subset enumeration.  
In choosing secure parameters for $f_{n,p,k}$, we must then choose $n$ such that 
the workload of an attack simply setting $c$ to be extremely small 
obeys desired security parameter $\lambda$.  We acknowledge 
that taking $c=\frac{(1+\epsilon)\log{n}}{\sqrt{n}}$ causes the algorithm to no 
longer run in time polynomial in $n$ because Alon's subset enumeration technique 
requires us to iterate through all size-$2\log_2{10/c}+2$ subsets of the vertex 
set; we do, however, note that doing so results in a potentially practical 
quasi-polynomial time attack when $p = \frac{1}{2}$.  In order for $f_{n,p,k}$ 
to satisfy security parameter $\lambda$, this must be taken into account.

$f_{n,p,k}$ embeds a clique in a graph drawn from $\mathcal{G}_{n,1/2}$
of size $(1+\epsilon) \log_2{n}$.  Thus, we need the spectral method using subset enumeration for $c=\frac{(1+\epsilon)\log_2{n}}{\sqrt{n}}$ to take at least $2^\lambda$ steps.  Note from 
section 4.1.3 that subset enumeration for this choice of $c$ must enumerate 
size-$s = 2\log_2{10/c}+2$ subsets.  Deriving this explicitly, we can give a lower bound 
on $s$:

\begin{align*}
s = 2\log_2{10/c}+2 &= 2\log{(10/(1+\epsilon)) \sqrt{n}} - 2\log{\log{n}} + 2 \\
& = \log{n} - 2\log\log{n} + 2 + \log_2{10/(1+\epsilon)} \\
\end{align*}

\noindent Using now the identity $\binom{n}{k} \geq (\frac{n}{k})^k$ and the 
fact that $\epsilon$ is between 0 and 1, we know that the number of subsets 
enumerated is bounded from below by

\begin{align*}
\binom{n}{s} & \geq (\frac{n}{s})^s \\
&= 2^{s(\log{n} - \log{s})} \\
&= 2^{\Omega(\log{n})(\log{n}-\log{\Omega(\log{n})})} \\
&= 2^{\Omega(\log^2{n})}
\end{align*}

\noindent We now need to enforce that the above quantity is greater 
than or equal to $2^\lambda$.  Given that 
$s = \log{n} - 2\log\log{n} + 2 + \log_2{5}$, the bounds used imply that we 
should take $n = 2^{\Omega(\sqrt{\lambda})}$. (This could perhaps be improved 
given a tighter analytic bound or expression for the binomial coefficient, 
but this will not make a difference in our final suggestions.  The choice of 
$n$ required to prevent the brute-force quasi-polynomial attack will 
be both identical and tight.)
\newline\newline
\noindent \textbf{Dekel's Method for Random Graphs } In section 4.1.4, we 
saw a probabilistic combinatorial method by Dekel which achieves the same bound 
as the spectral method of Alon in \cite{Spectral}.  This result is perhaps 
more significant with respect to our analysis, however, because it extends 
our capacity to find cliques of size $c \sqrt{n}$ in polynomial time to 
random graphs having arbitrary edge probabilities.  Noting the fact that 
$k \in [ \log_{1/p}{n}, 2\log_{1/p}{n} ]$ for $f_{n,p,k}$ and general $p$, 
we immediately have that the one-way function is vulnerable to Dekel's method 
when $p$ is large enough that $\sqrt{n}$ is near this range since $c$ will 
then either be close to or greater than $c_0$.

Furthermore, because Alon's subset enumeration technique may be applied to Dekel's method, the quasi-polynomial-time attack consisting of taking 
$c=\frac{(1+\epsilon)\log_{1/p}{n}}{\sqrt{n}}$ still applies.  In fact, the prospect 
of this attack makes the suggestion of increasing the edge probability from $\frac{1}{2}$
strictly worse in the case of Dekel's algorithm, as doing so strictly increases the 
efficiency of the attack.  To illustrate this point, consider the suggestion of 
Juels and Peinado in \cite{HidingCliques} to adjust $p$ such that the embedded clique 
would be of size $n^d$ for some fractional root $d$.  If $d > \frac{1}{2}$, then 
Dekel's method may be applied to invert $f$ in polynomial time.  If $d < \frac{1}{2}$, 
then $c = n^{d-1/2}$, and we therefore enumerate subsets of size 
$\log{n^{1/2-d}} = (1/2-d)\log{n}$: as $p$ (and therefore $d$) increases, 
the size of enumerated subsets (and consequently the time taken to complete the 
attack) decreases.  Thus, the suggestion we make in order to achieve security level $\lambda$ 
is to take $p$ as $\frac{1}{2}$ and set $n$ as we did in the case of Alon's spectral algorithm.
\newline\newline
\noindent \textbf{Sum-of-squares Relaxation } In section 4.1.5, we explored 
the concept of applying sum-of-squares (SOS) optimization in order to search
for planted cliques in random graphs.  We found, however, that there are somewhat
tight lower bound results in this area stating that this method is only 
able to find cliques of size $n^{1/2 - c}$ for variants which may be solved in 
polynomial time, and the extension of these bounds for programs able to be solved
in time $n^{O(d)}$ for general $d$ states that only cliques of size 
$n^{1/2-c(d/\log{n})^{1/2}}$ may be found.  In a manner similar to what we found 
for Alon's spectral method, we conclude that $f_{n,p,k}$ is vulnerable to 
SOS relaxation when $p=\frac{1}{2}$ and $k \in \Theta(n^{1/2-t})$ for an undetermined 
constant $t$.

A potential attack using SOS reduction involves choosing $d$ large enough 
that $n^{1/2-c(d/\log{n})^{1/2}}$ is equal to $2\log_2{n}$.  This, however, 
would require 
$d = \Omega(\log{n})$, meaning that the optimization procedure would 
take time $\geq n^{r_\epsilon\log{n}} = 2^{r_\epsilon\log^2{n}}$ to complete for some constant 
$r_\epsilon$.  In order to satisfy security parameter $\lambda$, we would need 
$2^{r_\epsilon\log^2{n}} \geq 2^\lambda$, meaning that $n \geq 2^{\sqrt{\lambda/r_\epsilon}}$.
If $r_\epsilon \geq 1$, we may simply take $n = 2^{\sqrt{\lambda}}$.
\newline\newline
\noindent \textbf{Feige's Algorithm } In section 4.2.1, we explored the 
best-known polynomial-time approximation algorithm for the maximum clique 
problem in general graphs by Feige.  Although this algorithm is considered 
to be the best we have, its approximation ratio is rather weak, only 
able to provide an approximation ratio of $O(n(\log{\log{n}})^2/\log^3{n})$.
In the case of random graphs drawn from $\mathcal{G}_{n,p}$, the maximum clique size
is w.h.p. $2\log_{1/p}{n}$, and in the case of the one-way function candidate 
$f_{n,p,k}$, $k$ is at least half of this size.  In general, then, we would 
require a generic approximation algorithm to require a ratio of at most 2 
to be useful for cryptanalysis.

They key result which allowed Feige to achieve this approximation ratio was a method to 
obtain a clique of size $(\log{n}/\log{\log{n}})^2$ in time 
$O(\binom{2jt}{t}n^c)$ for a graph having a clique of size $n/j$.  Since 
this guarantee is small, this method is only potentially pertinent for 
edge probabilities near $1/2$.  For such a random graph,
the embedded clique has size $(1+\epsilon)\log_2{n}$, so 
$j=\frac{n}{(1+\epsilon)\log{n}}$.  
It can be verified analytically that, for any $\epsilon$ and large enough $n$, the guarantee of  $(\log{n}/\log{\log{n}})^2$ is a guarantee of a clique with size equal to the one hidden 
in the graph.  We therefore need this method to require time $2^\lambda$ in order
for the security parameter to hold for $f_{n,1/2,k}$.

From Feige's analysis, one phase of this process takes time $\binom{2jt}{t}$ 
for $t = \frac{\log{n}}{\log\log{n}}$.  It would thus suffice for us to take
$n$ satisfying 

\begin{align*}
\binom{2jt}{t} &\geq (\frac{2jt}{t})^t \\
&\geq (\frac{n}{\log\log{n}})^{\log{n}/\log{\log{n}}} \\
&\geq 2^{\log^2{n}/\log{\log{n}} - \frac{\log{n}\log{\log{n}}}{\log{\log{n}}}} \\
&\geq 2^{\frac{\log^2{n}}{2\log{\log{n}}}} \\
&\geq 2^\lambda
\end{align*}

\noindent which holds if we take $n = 2^{\lambda^q}$ for some fractional 
root $q > \frac{1}{2}$.  Note that this 
suggestion comes from only a very loose analysis; it is perhaps possible to 
achieve security parameter $\lambda$ with smaller $n$.

\subsection{Consequences for the Suggested Primitives}

In this section, we tabulate recommendations for secure parameters for the function 
$f_{n,p,k}$ based upon vulnerabilities exposed by existing approximation algorithms.
We then use these recommendations to give a short discussion about the prospect 
of using $f_{n,p,k}$ and derived primitives in practice, suggesting future work in the 
way of exposing new attacks and extending the original premise motivating 
the suggestion of $f_{n,p,k}$ to obtain more viable primitives.

Below we tabulate and summarize our secure parameter suggestions conditioned on 
target security $\lambda$.  We denote by - that we have no recommendation.

\begin{figure}[H]
\centering
\begin{tabular}{ |c|c|c|c|  }
 \hline
 \multicolumn{4}{|c|}{\textbf{Secure Parameter Recommendations}} \\
 \hline
 \textbf{Adversary (Algorithm)}    & \textbf{Value of $n$} & \textbf{Value of $p$} & \textbf{Choice of $k$} \\
  \hline
 Brute Force  & $\begin{aligned}
 					\begin{cases}
 						2^{\Omega(\sqrt{\lambda})} & p \approx \frac{1}{2} \\
 						\lambda & p=2^{-2\log{\lambda}/\lambda} \\
 					\end{cases}
 				\end{aligned}$ & 
 				  $p \approx \frac{1}{2}$ or $p=2^{-2\log{\lambda}/\lambda}$ & $\epsilon \in [0,1]$ \\
 \hline
 Greedy & - & $p \geq \frac{1}{2}$ & $\epsilon$ closer to 1 \\
 \hline
 Metropolis(Worst Case) & $2^{\Omega(\sqrt{\log_{1/p}\lambda})}$ & $p \geq \frac{1}{2}$ & $\epsilon$ closer to 1 \\
 \hline
 Spectral (Alon) & $\begin{aligned}
 					\begin{cases}
 						2^{\Omega(\sqrt{\lambda})} & p \approx \frac{1}{2} \\
 						- & \text{else} \\
 					\end{cases}
 				\end{aligned}$ & $p \geq \frac{1}{2}$ & - \\
 \hline
 Dekel's Algorithm & $2^{\Omega(\sqrt{\lambda})}$ & $p = \frac{1}{2}$ & - \\
 \hline
 SOS Relaxation & $\begin{aligned}
 					\begin{cases}
 						2^{\Omega(\sqrt{\lambda/r_\epsilon})} & p \approx \frac{1}{2} \\
 						- & \text{else} \\
 					\end{cases}
 				\end{aligned}$ & $p \geq \frac{1}{2}$ & - \\
 \hline
 Feige's Algorithm & $\begin{aligned}
 					\begin{cases}
 						2^{\lambda^{q}}, q > 1/2 & p \approx \frac{1}{2} \\
 						- & \text{else} \\
 					\end{cases}
 				\end{aligned}$ & - & - \\
 \hline
 \textbf{ALL} & $2^{\lambda^q}$, $\frac{1}{2} << q \leq 1$ & $\frac{1}{2}$ & $\epsilon$ closer to 1\\
 \hline
\end{tabular}
\end{figure}

The final row in the above table is determined by taking the intersection of secure parameter 
ranges over all algorithms studied: this is precisely the set of parameters for which 
the one-way function of \cite{HidingCliques} is secure (for security parameter $\lambda$) against 
the approximation algorithms we are aware of.  To state what is shown in the table in 
plain English, the algorithms we've studied reveal that the candidate one-way function 
$f_{n,p,k}$ is, in fact, not secure in a reliable sense unless

\begin{itemize}
\item{The number of nodes in the graph $n$ is at least $2^{\lambda^q}$, where $q$ is 
between $\frac{1}{2}$ and 1.}
\item{The edge probability is $\frac{1}{2}$.}
\item{The $\epsilon$ determining the size of the planted clique is close to 1.}
\end{itemize}

\noindent This finding is especially interesting upon noticing that it directly contradicts 
the suggestion of the original authors in \cite{HidingCliques} to increase the edge 
probability to increase the complexity of brute-force attacks.  This specific finding, 
being derived from Dekel's algorithm \cite{Combinatoric}, largely stands to prove that recent work has 
actually uncovered something non-trivial about applications of the planted clique conjecture 
to cryptography, namely that such applications can be attacked if the edge probability is 
varied away from $\frac{1}{2}$.

Beyond revealing something about work completed so far, our exploration reveals areas 
where more work can be done as regards using the planted clique problem (and the 
maximum clique problem in general) to construct cryptographic primitives.  We offer a short 
discussion on these areas, separating them conceptually as \textit{offensive goals} and 
\textit{constructive goals}.

\begin{enumerate}
\item{\textit{Offensive goals } 
Our ability to (theoretically) attack the function candidate using only existing 
work in approximation invites consideration of other ways in which this manner of applying 
the planted clique problem may be attacked.}
\begin{enumerate}
\item{We saw that the primary contribution driving Feige's algorithm and the 
subset enumeration technique of Alon were able to drive the minimum-required value 
of the parameter $n$ to be exponential in some fractional power of $\lambda$.  A natural question 
to now ask is this: is it possible that we can find techniques (either new techniques or 
ways of applying existing ones) 
which drive this requirement even higher?  Could we, say, drive this minimum to be something 
like $2^{\Omega(\lambda)}$?  $2^{\Omega(\lambda^2)}$?

One particular route of inquiry: we saw that the Metropolis method proved only worst-case performance against 
random graphs.  An immediate question which comes to mind relates to how well this 
method performs in the average case or in the case where the method is supplied with partial 
information on the hidden clique.  Could modifications to the Metropolis method or 
other Monte Carlo methods drive higher the required minimum-secure graph size?
}
\end{enumerate}
\item{\textit{Constructive goals } Our exploration has revealed flaws in the construction of 
\cite{HidingCliques}.  While we have discussed the prospect of addressing the vulnerabilities 
we have found by simply increasing $n$ to be exponential in some function of $\lambda$ 
and fixing $p=\frac{1}{2}$, this introduces additional practical problems.  To illustrate, 
consider the output of $f_{n,p,k}$ when we take $\lambda$ as a common parameter, say 256.
According to our security recommendation, we would need to take $n\geq2^{\sqrt{256}} = 2^{16} = 65536$.
For $n=65536$ and edge probability $\frac{1}{2}$, we would have to store, transfer, and manipulate
graphs consisting of $65536$ nodes and $\binom{65536}{2}/2$ edges.  Even if we assume that 
each node and edge requires only a single bit of storage, such a graph would take up more than 130 megabytes of space.}
\begin{enumerate}
\item{If we wish to continue using the approach of \cite{HidingCliques}, we need to 
address the issue of the size of graphs stored.  One route of further exploration could be 
one which seeks a compressed method of representing random graphs containing planted cliques.
Towards making this idea concrete, we could choose some subset of $2^n$ graphs sampled from 
$\mathcal{G}_{2^{\lambda^q},1/2}$ each containing an embedded clique.  If there was a scheme to 
encode each of these graphs using $\lambda$ bits, we would solve the storage issue.  
Interestingly, if this scheme could somehow be designed incorporating 
information on the embedded clique as a 
trapdoor, it seems plausible that it might easily be transitioned to develop a public-key encryption 
scheme using the same assumption as the OWF.}
\item{One thought we must consider is the thought that, perhaps, the method of \cite{HidingCliques} 
will not yield a simultaneously secure and practical basis for cryptographic primitives.  Towards 
this end, perhaps we should go back to the drawing board: are the `hard' instances generated 
in the setting of the planted clique conjecture hard enough?  Perhaps there is another model of 
random graphs better suited to the task.}
\end{enumerate}
\end{enumerate} 

\section{Conclusion}

In this report, we have explored an effort \cite{HidingCliques} which attempts to give a 
cryptographic one-way function by relying on the hardness of the planted clique problem.  
Specifically, we have read, understood, and presented the details of the suggested construction, 
and we have formulated a manner in which to gauge the objective security it offers by 
scrutinizing its behavior when existing approximation algorithms are used as tools for 
cryptanalysis.  Upon completing our study, we found that our endeavor 
yielded three things: (1) a secure parameter guideline for the one-way function of \cite{HidingCliques} 
based upon a survey of our current attack capabilities, (2) direction for future work 
aimed at attacking the construction in \cite{HidingCliques}, and (3) insight into how we might 
improve such cryptographic applications of clique problems.

\bibliographystyle{acm}
\bibliography{proposal}

\end{document}